\documentclass[doublespacing]{elsart}

\usepackage{amssymb}

\usepackage{amsfonts}
\usepackage{amsmath}

\usepackage{bm}
\usepackage{epsfig}

\newcommand{\Fig}[1]{Fig.~(\ref{#1})}

\newcommand{\Table}[1]{Table~\ref{#1}}

\renewcommand{\vec}[1]{\bm{#1}}
\newcommand{\beq}{\begin{equation}}
\newcommand{\eeq}{\end{equation}}
\begin{document}

\begin{frontmatter}

\title{Configuration interaction study of single and double dipole plasmon
excitations in Na$_8$}

\author[l1,l2]{F. Catara}
\author[l1,l2]{D. Gambacurta} 
\author[l1,l2]{M. Grasso} 
\author[l2,l1]{M. Sambataro}

\address[l1]{Dipartimento di Fisica ed Astronomia, Via S. Sofia 64, 
I-95123 Catania, Italy}
\address[l2]{Istituto Nazionale di Fisica Nucleare, Sezione di Catania,
Via S. Sofia 64, I-95123 Catania, Italy}

\begin{abstract}
We carry out a microscopic analysis of the ground and excited states
of the Na$_8$ metal cluster within the jellium model. We perform a
series of
configuration interaction calculations on a Hartree-Fock basis and
construct
eigenstates of the Hamiltonian which carry up to 4-particle 4-hole
components.
Based on the analysis of the dipole transition strengths, we single out
those
states which can be interpreted as the collective dipole plasmon and
its double
excitations. These modes are found to possess a high degree of
harmonicity,
deviations from the harmonic limit remaining, however, of the
order of
10\%.
\end{abstract}

\end{frontmatter}

Collective vibrational states are known both in metal clusters
\cite{hab,ek} and
in nuclei \cite{bm}.
They are interpreted as the excitations of vibrational quanta, the so
called
plasmons or phonons. In particular, the dipole plasmon excitation in
clusters,
corresponding to an oscillation of the centre of mass of the valence
electrons
against that of the positive ions, is a very collective mode and
dominates the
response to a laser field. Microscopically, its frequency and strong
collectivity are quite well reproduced by the random phase approximation (RPA)
within the jellium
model. This model \cite{gj,mgj} considers a uniform positive charge
distribution
generated by
the ions (the jellium) which interacts with a cloud of delocalized valence
electrons
(also interacting among themselves) via the Coulomb interaction.
RPA predicts that the plasmon oscillation is
perfectly harmonic \cite{schu}, i.e. that states corresponding to a n-fold
excitation of
the plasmon exist and their energy is equal to $n$ times that of the single
plasmon. If the jellium-electron interaction can be approximated by a
harmonic
oscillator potential, as it is reasonably true for a highly positively
ionized cluster, this is what one expects on general grounds \cite{ycg}.
Indeed, in
that
case the total electron eigenfunctions are simply products of the type
\begin{equation}
\Psi (\vec{r}_1,\vec{r}_2,...\vec{r}_N)=\psi_{nl}(\vec{R})\phi_{\nu \lambda}
(\vec{r}'_1,\vec{r}'_2,...\vec{r}'_N)~,
\label{1}
\end{equation}
where $\vec{R}$ is the coordinate of the centre of mass of the electrons
with
the harmonic oscillator quantum numbers $(nl)$ describing its motion, while
${\vec{r}'_i}$ are the intrinsic coordinates of the electrons with $\nu$ and
$\lambda$ classifying the harmonic oscillator states of their relative
motion.
A dipole external field acting on the ground state of the system, $\psi_{00}
\phi_{00}$, causes the transition to $\psi_{01} \phi_{00}$, acting on the
latter excites the states  $\psi_{10} \phi_{00}$ and  $\psi_{02} \phi_{00}$,
and so on, generating a perfectly harmonic band. Of course, other bands
exist,
based on different intrinsic motion states $\phi_{\nu \lambda}$, with no
electromagnetic transitions among them. This was checked numerically in Ref.
\cite{casa} for the very simple case of two interacting electrons moving in
a
harmonic oscillator potential. When this potential is replaced by the Coulomb
potential generated by the jellium charge distribution the above level
scheme
is modified and it was found in Ref. \cite{casa} that different bands still
exist,
deviating from the harmonic limit and with non zero multipole transition
probabilities among them. These results show that the anharmonicities in the
dipole plasmon excitation are due to the coupling between the intrinsic and
centre of mass motions of the electrons.

This has been recently discussed quite
in general in Ref. \cite{guet}. 
In this work, the authors make explicit in a very clear way
the coupling between intrinsic and centre of mass motions
by considering an expansion of the ionic
background potential $V_{ion}(\vec{r}_i)=V_{ion}(\vec{r}'_i+\vec{R})$ in
power
series with respect to $\vec{R}$.
Appreciable deviations from the harmonic scheme are found within this
approach
\cite{guet}. For example, for a Na$_{93}^+$ cluster it is found that the
most
important optical dipole transitions are separated by a $\Delta E_n$ equal
to
2.8 $eV$, 3.07 $eV$ and 3.26 $eV$ for the single ($n=1$), double ($n=2$)
and triple
($n=3$) plasmon excitations, respectively. These results should be
compared with the
contradictory ones found in Refs. \cite{hagino} and \cite{noi},
namely essentially
zero and huge anharmonicities, respectively. This is due to the fact that the
coupling between intrinsic and centre of mass motions is almost completely
neglected in Ref. \cite{hagino}. On the other hand, the boson expansion used
in Ref.
\cite{noi} was truncated at the same order as previously done for the study
of the double excitation of Giant Resonances in atomic nuclei \cite{noinuclei}.
In the latter case, anharmonicities of the order of a few hundred keV as
compared with the harmonic limit energy of $20 \div 30 MeV$ were found.
The huge
anharmonicities found in metallic clusters are probably an indication that the
convergence of the boson expansion is much slower in that case and this can be
related to the long range of the Coulomb interaction \cite{ycg}. To our
knowledge, there is no experimental clear evidence on the existence of
states corresponding to the double excitation of the dipole plasmon
\cite{hab1,hab2}. On the other hand, in atomic nuclei, the existence of 
multiphonon states has been known for many years \cite{bm} and the 
anharmonicities in their excitation spectra have been found to play a role in 
several physical processes \cite{hara}.

This and the very interesting results found in Ref. \cite{guet}
encouraged us to attempt a configuration
interaction (CI) calculation for small metal clusters. Indeed, this is a
 priori the best method to obtain the energy spectrum. On the other hand,
since the states we want to study are quite high in energy, namely at about
twice or more that of the plasmon, the configuration space required to get
reliable solutions becomes very rapidly prohibitively large when the size of
the cluster increases. We have thus limited our analysis to Na$_8$.
Similar calculations for several light Na clusters have been reported in
Ref. \cite{fin} where, however,
only the ground state and the singly excited dipole
state were studied.

Let us then shortly describe our calculations and discuss
the results. Within
the jellium model the motion of the valence electrons is determined by the
hamiltonian
\begin{equation}
H=\sum_i h_i + \sum_{i<j} v_{ij}~,
\label{2}
\end{equation}
with
\begin{equation}
h_i=- \frac{\hbar^2}{2m} \nabla^2_i + V(r_i)~; \; v_{ij}=\frac{e^2}{4\pi
\epsilon_0} \frac{1}{|\vec{r}_i-\vec{r}_j|}~,
\label{3}
\end{equation}
and
\begin{equation}
V(r)=\frac{Ze^2}{4\pi\epsilon_0}
\Biggl\{\begin{array}{c}
(1/2r_c)(r^2/r_c^2-3)~for~r\leq r_c\\
~~~~~~~~~~~~~~~~~~~~~~~~~~~~~~~~~~~~~~~~~~~~~~~,\\
-1/r~~~~~~~~~~~~~~~~~~for~r\geq r_c
\end{array}
\label{4}
\end{equation}
where $r_c$ is the radius of the jellium sphere, i.e. $r_c=r_sN_{e}^{1/3}$ with
$r_s$ the Wigner-Seitz radius which is 4 a.u. 
for Na and $N_{e}$ is the number of electrons. We are aware of the fact that, 
for a small metal cluster, the jellium approximation may be not 
completely adequate for a quantitative comparison with experimental data. 
However, it already contains many important physical features while 
allowing not too heavy calculations as compared with more elaborated 
models. 

As a first step we make a Hartree-Fock (HF) calculation in order to fix
single particle ($sp$) energies and wavefunctions. This is done by allowing the
wavefunctions to be superpositions of harmonic oscillator 
wavefunctions (h.o.w.f.'s).
The number of h.o.w.f.'s has to be chosen large enough to get a HF ground
state
energy satisfactorily stable for small variations of the h.o. parameter
around
the value giving the minimum.
Next, we construct all Slater determinants with fixed values of the
projection
$M_L$ of the total angular momentum, of the projection $M_S$ of the total
spin
and of the parity $\pi$. This set is truncated by
i) truncating the $sp$ basis;
ii) putting a maximum value $n$ for the number of particle-hole 
excitations ($np$-$nh$) with 
respect to the HF ground state and/or for the unperturbed excitation energy.
We have considered several truncations corresponding to
i) up to 10 HF orbitals above the Fermi level;
ii) all Slater determinants having up to $3p$-$3h$ configurations
and containing those  $4p$-$4h$ configurations whose unperturbed energy
with respect
to the
HF ground state is less than a given cutoff energy $E_c$. The use of the HF
basis should optimize the convergence of the results because part of the
correlations are already taken into account in the reference state.

The largest basis we have been able to manage has dimension $\sim$ 700000;
it is to be noted, however, that many matrix elements of the hamiltonian in
such a basis are zero: the hamiltonian matrix is sparse. We have then
used
a $NAG$ library routine especially intended for such a case.
 The routine finds the $N$
eigenvalues of largest absolute value and the corresponding eigenvectors.
This method is very suitable for our case since we are interested in the 
lowest negative eigenvalues which are the largest ones in absolute value.
The time required by the routine to find the solutions increases very rapidly
with $N$.
Therefore, $N$ has to be taken as low as possible. From RPA we know that the
excitation energy of the dipole plasmon is located around 3 $eV$. Since we are
interested in single and double dipole plasmon excitations, we have to choose
$N$ such that all eigenvalues up to $\sim$ 7 $eV$ are determined,
and this would
mean a quite large $N$ since the states of the basis we use do not
have a definite
value of the total angular momentum and spin. On the other hand, since the
hamiltonian commutes with $\hat{\vec{L}}$ and $\hat{\vec{S}}$, its
eigenvectors are also eigenvectors of $\hat{\vec{L}}^2$ and $\hat{\vec{S}}^2$
in addition to $\hat{L}_z$, $\hat{S}_z$ and parity. 
We take advantage of this and use the following procedure 
to select the eigenstates
of $H$ belonging to some definite value of the angular momentum $\bar{L}$.
By running the calculation with $M_{L}=\bar{L}$, 
the eigenstates with $L<\bar{L}$ are trivially eliminated. 
By adding to the hamiltonian a term 
 $\alpha [\hat{\vec{L}}^2-\bar{L}(\bar{L}+1)]$ with $\alpha>0$
 the eigenvalues corresponding to $L>\bar{L}$ are shifted up. Therefore, 
 if $\alpha$ is chosen large enough, only those with 
$L=\bar{L}$ are selected as
 the lowest ones (i.e. with the largest absolute values). 
In reality one must be careful since, if $\alpha$  is chosen very large, 
 the eigenvalues corresponding to $L \gg \bar{L}$ are so 
much shifted up that they become positive and 
 the  largest ones. Therefore, $\alpha$  has to be kept not too large 
and then a few eigenvalues not corresponding
  to $\bar{L}$ may be mixed in the region of the spectrum we look at.
  However, this problem is easily eliminated by calculating after the 
diagonalization 
  the angular momentum of each eigenstate. The same reasoning applies 
also for the spin
  and we add to the hamiltonian another term
  $\beta [\hat{\vec{S}}^2-\bar{S}(\bar{S}+1) ]$ with $\beta>0$.
  The above sketched procedure is very effective and it has allowed us to 
limit the number of eigenvalues we are interested
   in within a maximum value of 50, for each of the three cases relevant 
for our scopes, namely $L^{\pi}S = 1^- 0, 0^+ 0$ and $2^+ 0$,
  and an excitation energy less than 7 $eV$. 
The energies of the lowest 12 HF $sp$ states are reported in
\Table{tab1}.

Each $sp$ state has been expressed as superposition of h.o.w.f.'s,
with principal quantum number running from 0 to 8. The HF states are labeled
by the principal
quantum number $n$ of the predominant component and the angular quantum
number
$l$.

For a start, we have performed two series of
calculations by considering those $sp$ states and including all Slater
determinants with up to  $2p$-$2h$
and  $3p$-$3h$ configurations. By comparing the two series of results 
we have observed that
the  $2p$-$2h$ space is far from being sufficient for a good description of the
states we are interested in.
Indeed,
the inclusion of the  $3p$-$3h$ configurations strongly modifies the
energies of the excited states and, to a less extent, of the ground state (see 
Table II).
In order to go further one has to include  $4p$-$4h$ configurations. However,
in this case the number of configurations is too large and it is necessary to 
introduce an energy cutoff $E_{c}$, i.e. to include only those  $
4p$-$4h$ Slater determinants
 whose unperturbed energies are not higher than $E_{c}$ above the $HF$ 
ground state.
 By repeating the calculations with $E_{c}$ increasing from $E_{c}= 19~ eV$ 
to $E_{c}= 25 ~eV$
 we get a lower and lower ground state energy as shown in \Table{tab2}.
 Looking at the third column of the table, where we show the difference 
between the ground state energies relative to
 two successive calculations,  one can conclude that a very good 
 numerical convergence has been reached for the ground state energy. Indeed, 
the values
  obtained with $E_{c}=24$ and 25 $eV$ differ only by 2 $meV$. 
However, we are also interested in
  excited states having a quite high energy and convergence must be checked 
also for them. 
  In order to do that we have calculated the root mean square 
value $ \sigma=\sqrt{ 1/N \sum_{i=1}^{N}\Delta_{i}^{2}}$
  of the shifts  $\Delta_{i}$ obtained in two calculations with $E_{c}$  
differing by 1 $eV$.
  By considering all states having excitation energy less than 7 $eV$, 
for the case 
  $L^{\pi}S =  0^+ 0$, we have got $\sigma=$ 0.064 $eV$ when going 
from $E_{c}=$ 20 $eV$ to 
  $E_{c}=$ 21 $eV$ while   $\sigma=$ 0.039 $eV$ in going from $E_{c}=$ 24 
$eV$ to 
  $E_{c}=$ 25 $eV$. It is also to be noted that, in the latter case, the 
largest shift is $\Delta =$ 0.046 $eV$ 
  corresponding to 0.7 \%. Similar results have been obtained for the 
$2^+ S=0$ and $1^-S=0$ spectra.
  
 A further comment has to be added. 
  In order to be able to increase so much the energy cutoff for  $4p$-$4h$ 
 configurations we have followed a suggestion of Ref. \cite{fin}. Namely, 
the high
 angular momentum $0g$ and $0h$ single particle states have been suppressed 
from the basis.
 We have checked that, for $E_{c}=$ 20 $eV$, the calculations with and 
without those states
 give almost indistinguishable results, more precisely a maximum shift 
 of 0.2 \% in the considered energy region. The next two $sp$ states
  above the twelve ones we have considered in the basis
  have $l$=6 and $l$=7. Therefore,  their inclusion should not modify the 
spectra.
  Still above there is a $2d$ orbital, but  its $HF$ energy  is 2.19  $eV$
  and we can reasonably neglect its contribution.
  From the above considerations we conclude that a satisfactory convergence 
has been reached and the
  so obtained energies and wavefunctions can be reliably used in the analysis 
of anharmonicities.

\begin{figure}
\begin{center}
\epsfig{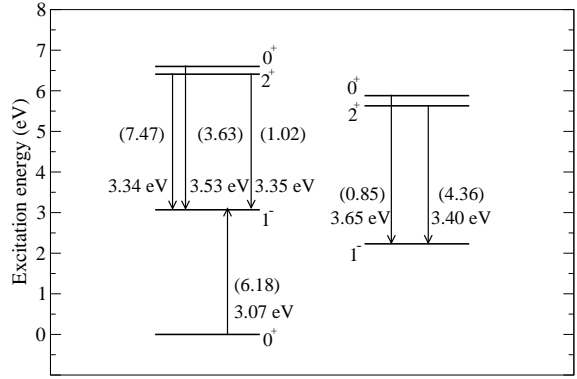}
\end{center}
\caption{The largest $|T_{fi}(E1)|^{2}$ from the 0$^+$ $S = 0$ and 2$^+$ 
$S = 0$ states 
 to the 1$^-$ $S = 0$ ones.  The values, in $\AA^2 $, are indicated in 
brackets.
 The calculation is done within the to  $4p$-$4h$ space  with an energy cutoff 
equal
to 25 $eV$ and 12 HF orbitals. The states are grouped in two bands based on 
the 
ground state and on the lowest 1$^-$ $S = 0$ state. For the excited states
we report the  energies corresponding to the transitions indicated by the 
arrows.
The two arrows in correspondence with the highest $2^{+}$ state refer to two 
almost
degenerate states which are not distinguishable in the figure (see the text).
\label{fig1}}
\end{figure}

In order to single out the states which can be interpreted as the collective
 dipole plasmon and its double excitations we have calculated the electric
 dipole transition strengths from the initial states
 $|\psi_{i}>$, 0$^+$ $S = 0$ and  2$^+$ $S = 0$,   to the final states 
$|\psi_{f}>$, 
 1$^-$ $S = 0$. More precisely, we have calculated the 
 following quantity:
 \begin{equation}\label{eq:matrix}
 |T_{fi}(E1)|^{2}=
 |\sum_{\alpha \alpha'}t_{\alpha
 \alpha'}^{(E1)}
 <\psi_{f}|a_{\alpha}^{\dag}a_{\alpha'}|\psi_{i}> |^{2}~,
 \end{equation}
 with $\alpha=(n_{\alpha},l_{\alpha},\sigma_{\alpha})$
 and
 \begin{equation}\label{eq:red_max}
t_{\alpha \alpha'}^{(E1)}=(\alpha \parallel Y_{1}\parallel
\alpha')=R_{n_{\alpha}l_{\alpha},n_{\alpha'}l_{\alpha'}}
(l_{\alpha}\parallel Y_{1}\parallel
l_{\alpha'})\delta_{\sigma_{\alpha}\sigma_{\alpha'}}~,
\end{equation}
\begin{equation}\label{eq:int}
R_{nl,n'l'}=\int
r^{3}\varphi^{\ast}_{nl}(r)\varphi_{n'l'}(r)dr~.
\end{equation}
The $(l\parallel Y_{\lambda}\parallel l')$ in eq. \eqref{eq:red_max} are 
the reduced matrix elements 
defined as in Ref. \cite{hara} and 
$\varphi_{nl}$ in eq. \eqref{eq:int} are the Hartree - Fock $sp$ wave 
functions. 
 In \Fig{fig1} we report the largest $|T_{fi}(E1)|^{2}$ values (greater than 
0.8 $\AA^2$) obtained in the calculation performed with
 12 $HF$ states and within the  $4p$-$4h$ space with $E_{c}$ =25 $eV$.  
As we are interested in the study of the collective dipole plasmon and its 
double excitations, we show only the transitions involving the $1^-$ states 
located around the single plasmon excitation energy.

The selected levels can be grouped in two ``bands'' based on the ground state 
and
on the lowest 1$^-$ $S = 0$ state ($1_{1}^{-}$), respectively. The two bands 
are
essentially not connected by dipole transitions. The second 1$^-$
$S = 0$ state ($1_{2}^{-}$) at 3.07 $eV$ excitation energy is strongly coupled
to the ground state and can be identified with the dipole plasmon
excitation. The only three excited states having a large dipole
transition strength to this $(1_{2}^{-})$ state are two 2$^+$ $S =
0$ states, at 6.41 and 6.42 $eV$, and one 0$^+$ $S = 0$ state at 6.60 $eV$,
corresponding to an energy jump of 3.34, 3.35 $eV$ and 3.53
$eV$ respectively, to be compared with the 3.07 $eV$ excitation energy of 
the ($1_{2}^{-}$) state.
 They might be identified as corresponding to
 double excitations of the dipole plasmon. Indeed, they are very
close to each other and their energy is not far from twice that of
the single plasmon. The deviations from the harmonic limit are of
the order of  10\%. As can be seen from the figure, the strength of the 
transition connecting the $(1_{2}^{-})$ state to the two-plasmon states is 
12.12 $\AA^2$ to be compared with the harmonic value of 12.36 
$\AA^2$ corresponding to the double of the one-plasmon strength. We also 
note that this strength is fragmented between the  2$^+$ $S = 0$ and
0$^+$ $S = 0$ states in a proportion close to the harmonic limit, 
i.e. 4/3 and 2/3 respectively.

The 2$^+$ $S = 0$ and
0$^+$ $S = 0$ having a strong transition to the ($1_{1}^{-}$) state (see  
the ``lateral band'' in
the figure) lie at 3.40 $eV$ and 3.65 $eV$ above it. This makes
plausible their interpretation as  collective dipole excitations
built on top of the $(1_{1}^{-})$ state, all these states being 
characterized by an intrinsic motion wave function $\phi_{01}$ 
(see eq. (1)). This also justifies the extremely small value 
(0.002 $\AA^2$) of the transition strength between the ground state 
and the  $(1_{1}^{-})$ state.

The quality of the results obtained in the present calculations can be judged
by looking at sum rules. It is well known \cite{schu} that, if $|0\rangle$ and
$|\nu\rangle$
are the exact ground and excited states of a system, then the following
equality
holds
\begin{equation}
\sum_{\nu} (E_{\nu} -E_0)|\langle\nu |T_{\lambda} |0\rangle |^2=
\frac{1}{2}\langle\nu |[T_{\lambda} ,[H,T_{\lambda} ]]|0\rangle ~,
\label{5}
\end{equation}
where $H$ is the total hamiltonian and $T_{\lambda}$ the transition
operator of multipolarity $\lambda$.
The r.h.s. of the above equation can be evaluated
exactly and is (see eq. (2.47) of Ref. \cite{hara})
\begin{equation}
EWSR=\frac{\hbar^2\lambda(2\lambda +1)^2}{8\pi m}N_{e}\langle r^{2\lambda -2}
\rangle ~,
\end{equation}
where $m$ is the mass  of the electron 
and $\langle r^{2\lambda -2}\rangle$ is the expectation value of the indicated
quantity in the ground state. Since we are looking at $\lambda=1$ transitions,
the EWSR is completely independent of the ground state and is equal to 21.83
$\AA^2 \cdot eV$.
The l.h.s. of eq. (\ref{5}) turns out to be 21.62 $\AA^2 \cdot eV$  which is 
99.04\%
of EWSR. Therefore, we can conclude that our calculated energies and 
wavefunctions satisfy
up to an extremely good level this very stringent condition.
In order to further check the numerical convergence we have compared the 
results obtained
with 12 $HF$ orbitals and cutoff energies equal to $E_{c}=$ 22, 23, 24, 25 
$eV$,
finding a smaller and smaller variation at each step, the last one being 
0.06 $\AA^2 \cdot eV$, i.e. 0.2 \%.
It is also worth mentioning that the
contribution of the second $1^-$ state to the sum is 19.01 $\AA^2 \cdot eV$,
which
means 87\% of the total. Therefore, its identification with the collective
dipole plasmon state is very well justified. \\
By summarizing, in this work we have
carried out a microscopic analysis of the ground and excited states
of the Na$_8$ metal cluster within the jellium model. We have performed
 a series of
configuration interaction calculations in a Hartree-Fock basis and
constructed
eigenstates of the Hamiltonian which carry up to 4-particle 4-hole
components.
Based on the analysis of the dipole transition strengths,
we have singled
 out those
states which can be interpreted as the collective dipole plasmon and
its double
excitations. These modes have been found to possess a high degree of
harmonicity,
deviations from the harmonic limit remaining, however, of the order of
10\%. These values are consistent with the anharmonicities found by
Gerchicov et al. \cite{guet}, using a different technique, in heavier
clusters. We want to stress that in principle a configuration 
interaction calculation
is the most accurate approach to reproduce the spectrum of the system and then 
to evaluate its anharmonicities.
On the other side, a limitation of this kind  of calculations is that
they are very heavy numerically and this is the reason why
we  limited our analysis to the small cluster Na$_8$.

\newpage

\begin{table}
\begin{center}
\begin{tabular}{|c|c|c|c|c|c|c|}
\hline
$\vec{n \;l}$ & $\vec{0 \;0}$ & $\vec{0 \;1}$& $\vec{0 \;2}$ &
$\vec{1 \;0}$ & $\vec{1 \;1}$ & $\vec{1 \;2}$ \\
\hline
\hline
\textbf{E (eV)} & -6.85 & -4.38 &  0.15 &  0.25 & 0.50 & 0.83 \\
\hline
$\vec{n \;l}$ & $\vec{0 \;3}$&
$\vec{2 \;0}$ & $\vec{0 \;4}$ & $\vec{0 \;5}$ & $\vec{2 \;1}$ &
$\vec{1 \;3}$      \\
\hline
\hline
\textbf{E (eV)} & 0.86 & 1.12 & 1.16 & 1.46 & 1.62 &  1.76  \\
\hline
\end{tabular}
\end{center}
\caption{Energies ($eV$) of the lowest 12 HF $sp$ states; $n$ is the principal
quantum number of the predominant component and $l$ is the angular
quantum number.}
\label{tab1}
\end{table}

\begin{table}
\begin{center}
\begin{tabular}{|c|c|c|c|c|c|c|}
\hline

~~~& $E_{0} ~~(eV)$ & $\Delta_{0}~~ (eV)$\\
\hline
 $2p$-$2h$ & -142.095 & -\\
 \hline
 $3p$-$3h$ & -142.176 & -0.081\\
 \hline
 $4p$-$4h, E_{c}=19~ eV$ & -142.235 & -0.059\\
 \hline
 $4p$-$4h, E_{c}=20~ eV$ & -142.248 & -0.013\\
 \hline
  $4p$-$4h, E_{c}=21~ eV$ & -142.252 & -0.004\\
 \hline
  $4p$-$4h, E_{c}=22~ eV$ & -142.258 & -0.006\\
 \hline
  $4p$-$4h, E_{c}=23~ eV$ & -142.275 & -0.017\\
 \hline
  $4p$-$4h, E_{c}=24~ eV$ & -142.283 & -0.008\\
 \hline
  $4p$-$4h, E_{c}=25~ eV$ & -142.285 & -0.002\\
\hline
\end{tabular}
\end{center}
\caption{In the second column the ground state energies are reported for 
each calculation.
In the third column $\Delta_{0}$ represents the difference between the 
corresponding ground state
 energy and that reported in the previous line. }
\label{tab2}
\end{table}

\end{document}